\documentclass[12pt]{amsart}
\usepackage[latin9]{inputenc}
\usepackage{bm}
\usepackage{amsmath}
\usepackage{amsthm}
\usepackage{amssymb}
\usepackage[authoryear]{natbib}
\usepackage[unicode=true,
 bookmarks=false,
 breaklinks=false,pdfborder={0 0 1},backref=section,colorlinks=false]
 {hyperref}
\usepackage{breakurl}

\usepackage[usenames,dvipsnames,svgnames,table]{xcolor}

\usepackage{fourier}
\hypersetup{colorlinks=true, urlcolor=MidnightBlue, citecolor=MidnightBlue, linkcolor=MidnightBlue}

\makeatletter
\theoremstyle{plain}
\newtheorem{thm}{\protect\theoremname}
  \theoremstyle{plain}
  \newtheorem{fact}{\protect\factname}
  \theoremstyle{definition}
  \newtheorem{defn}{\protect\definitionname}
 \theoremstyle{definition}
 \newtheorem*{defn*}{\protect\definitionname}
  \theoremstyle{plain}
  \newtheorem*{thm*}{\protect\theoremname}
  \theoremstyle{plain}
  \newtheorem{prop}{\protect\propositionname}

\usepackage{amsfonts}
\usepackage{bm}
\usepackage{etex}
\usepackage{setspace}

\usepackage{graphics}
\usepackage{epsfig}\usepackage{verbatim}\usepackage{bm}\usepackage{latexsym}\usepackage{url}\usepackage{rotating}

\usepackage{mathrsfs}
\usepackage{multirow}
\usepackage{array}
\usepackage{url}

\makeatother

  \providecommand{\definitionname}{Definition}
  \providecommand{\factname}{Fact}
  \providecommand{\propositionname}{Proposition}
  \providecommand{\theoremname}{Theorem}
\providecommand{\theoremname}{Theorem}

\begin{document}

\title[Linear Network Games]{Introduction to Network Games \\ with Linear Best Responses}

\author{Benjamin Golub}
\date{May 2018}

\begin{abstract}
	This note, suitable for a lecture in an advanced undergraduate or basic graduate course on the economic theory of networks, exposits basic ideas of linear best-response games and their equilibria.
\end{abstract}

\maketitle

\section{Basic setup and results}

We study the game of \cite*{Ballester2006}, who introduced the idea of the connection between Nash equilibria of a certain kind of game and centrality measures that we will derive is due to them.

Consider a complete-information game where each player (also called
agent) $i\in N=\{1,2,\ldots,n\}$ simultaneously selects a real-valued
action $a_{i}\geq0$ and receives a real-valued payoff $u_{i}(a_{1},a_{2},\ldots,a_{n})$
that depends on everyone's action. Suppose that each agent $i$'s
best-response function is given by 
\begin{equation}
\text{BR}_{i}(\bm{a}_{-i})=\alpha\sum_{j}w_{ij}a_{j}+b_{i}.\label{eq:BR}
\end{equation}
Here 
\[
\alpha>0,\quad\bm{W}=(w_{ij})_{i,j\in N},\text{ and }(b_{i})_{i\in N}
\]
are constants\textemdash parameters of the model that do not depend
on $\bm{a}=(a_{i})_{i\in N}$. The matrix $\bm{W}$ is irreducible,\footnote{We call a matrix irreducible if the corresponding weighted, directed
graph is strongly connected. A $1$-by-$1$ nonnegative matrix is
said to be irreducible if its sole entry is positive.} with $W_{ii}=0$ for every $i$, and all its entries are nonnegative.
All the $b_{i}$ are positive. 

\subsection{Existence result\label{subsec:existence}}

Recall that $r(\bm{A})$ is the spectral radius of a matrix $\bm{A}$,
which has two equivalent definitions: (i) the maximum absolute value
of any eigenvalue; and (b) the definition you studied in Problem Set
1, Problem 4.
\begin{thm}
\label{thm:eqm-characterization}If $r(\alpha\bm{W})<1$ then there
is exactly one pure-strategy Nash equilibrium of the game described
above, given by 
\begin{align*}
\bm{a}^{*} & =(\bm{I}-\alpha\bm{W})^{-1}\bm{b}\\
 & =\sum_{\ell=0}^{\infty}\alpha^{\ell}\bm{W}^{\ell}\bm{b}.
\end{align*}
\end{thm}
The result can be established by manipulating the assumed best responses
of each agent to show that as long as $(\bm{I}-\alpha\bm{W})^{-1}$
is well-defined, then 
\[
\bm{a}^{*}=(\bm{I}-\alpha\bm{W})^{-1}\bm{b}.
\]
It is left as an exercise to show that $(\bm{I}-\alpha\bm{W})^{-1}$
exists, has the claimed Neumann series expansion, and is nonnegative.\footnote{Let $\bm{A}$ be an irreducible $n$-by-$n$ matrix whose entries are all nonnegative. (a)  Assume that $\sum_{k=K}^{\infty} \bm{A}^k$ tends to the all-zeros matrix entrywise as $K\to \infty$.	Show that $$(\bm{I}-\bm{A})^{-1} = \sum_{k=0}^{\infty} \bm{A}^k,$$ where $\bm{I}$ is an appropriately-sized identity matrix. That is, show that the expression $ \sum_{k=0}^{\infty} \bm{A}^k$ is well-defined and that it is the inverse of the matrix $\bm{I}-\bm{A}$.	(b) Conclude that all the entries of $(\bm{I}-\bm{A})^{-1} $ are nonnegative. (c) Using the Perron-Frobenius Theorem, show that the assumption (first sentence) of (a) holds if the spectral radius of $\bm{A}$ is strictly less than $1$.}

\subsection{Interpreting the entries of $\bm{W}^{\ell}$}

For any walk $s=(i(1),i(2),\ldots,i(\ell+1))$ define its \emph{weight}
to be the product of the weights of the edges constituting that walk:
\[
\lambda(s)=w_{i(1)i(2)}w_{i(2)i(3)}\cdots w_{i(\ell)i(\ell+1)}.
\]
Let $\mathcal{W}^{\ell}(\bm{W};i,j)$ be the set of all walks of length
$\ell$ from $i$ to $j$.
\begin{fact}
The following identity holds: for every positive integer $\ell$,

\[
(\bm{W}^{\ell})_{ij}=\sum_{s\in\mathcal{W}^{\ell}(\bm{W};i,j)}\lambda(s).
\]
\end{fact}
That is, the $(i,j)$ entry of $\bm{W}^{\ell}$ is the sum of the
weights of walks of length $\ell$ from $i$ to $j$. Think about
the special case of $\bm{W}=\bm{G}$. This boils down to counting
walks.

What do these walks correspond to game-theoretically? In view of the
formula $\bm{a}^{*}=\sum_{\ell=0}^{\infty}\alpha^{\ell}\bm{W}^{\ell}\bm{b}$,
the indirect effect of perturbations to $b_{i}$ on the action $a_{j}$
via chains of best responses of length $\ell$. For a different interpretation of the walks, in terms of a process of iterated best-responses, see \cite[Section 3.1.3]{golub-sadler}.

\section{Bonacich centrality}
\begin{defn}
Let $\widehat{\bm{W}}$ be a nonnegative irreducible matrix, $\widehat{\alpha}\in(0,r(\bm{\widehat{\bm{W}}})^{-1})$
be a real number, and $\widehat{\bm{b}}$ a nonnegative \emph{base
vector}. Then the vector of \emph{Bonacich centralities} in $\widehat{\bm{W}}$
with parameter $\widehat{\alpha}$ and base vector $\widehat{\bm{b}}$
is defined to be 
\[
\bm{\beta}(\bm{\widehat{\bm{W}}};\widehat{\alpha},\bm{\widehat{\bm{b}}})=(\bm{I}-\alpha\bm{\widehat{\bm{W}}})^{-1}\widehat{\boldsymbol{b}}.
\]
The ``default'' value of $\bm{b}$ is $\bm{1}$, the vector of all
ones.
\end{defn}
We are using hats to distinguish these from the parameters of our
game, since the arguments of $\bm{\beta}$ can be anything (though
of course we will apply it to the game shortly). Note that the Bonacich
centrality satisfies 
\[
\bm{\beta}=\bm{\widehat{\bm{b}}}+\widehat{\alpha}\bm{\widehat{\bm{W}}}\bm{\beta}.
\]

The idea behind this equation is that $\beta_{i}$ is a measure of
$i$'s network centrality. It is the sum of a ``base'' level $\widehat{b}_{i}$
and a socially derived part $\widehat{\alpha}\sum_{j}\widehat{W}_{ij}\beta_{j}$,
where $\widehat{W}_{ij}$ describes the part of $\beta_{j}$ that
accrues to $i$. The natural interpretation in many cases is that
$\widehat{W}_{ij}$ is the amount of attention that $i$ gets from
$j$, so is naturally interpreted as a link from \emph{$j$ to $i$.}

The reference for this notion is \cite{bonacich}.
However, Leontief studied very similar ``centrality'' measures around
30 years before that, and the Leontief inverse, $(\bm{I}-\alpha\bm{W})^{-1}$, comes up in the current study of production networks \citep{acemoglu2012network,baqaee2017macroeconomic}.

Now, we know from the Neumann series that
\[
\bm{\beta}(\bm{\widehat{\bm{W}}};\widehat{\alpha},\bm{\widehat{\bm{b}}})=\sum_{\ell=1}^{\infty}\alpha^{\ell}\bm{\widehat{\bm{W}}}^{\ell}\bm{\bm{\widehat{\bm{b}}}}.
\]
Thus we can interpret an agent's Bonacich centrality in terms of sums
of walk weights. You did a similar thing for unweighted walks in Problem
Set 1.

In our applications, we will be interested in the Bonacich centrality
vector, $\bm{\beta}(\bm{\widehat{\bm{W}}};\widehat{\alpha},\bm{\widehat{\bm{b}}})$,
for various matrices $\bm{\widehat{\bm{W}}}$ and various vectors
$\widehat{\bm{b}}$. For example, clearly in our game
as set up in the beginning of Section \ref{subsec:existence} we may
write the unique equilibrium identified in Theorem \ref{thm:eqm-characterization}
as
\[
\bm{a}^{*}=\bm{\beta}(\bm{W};\alpha,\bm{b}).
\]

We can also describe another important aspect of our game via a different
application of Bonacich centrality, i.e., with different parameters.
Define the \emph{total activity level} by $A^{*}=\sum_{i}a_{i}^{*}$
and define the \emph{keyness} of $i$ to be 
\[
k_{i}=\frac{dA^{*}}{db_{i}}.
\]
This is the amount by which an exogenous change in $b_{i}$ affects
the \emph{aggregate equilibrium action} in the game. (Though we have
defined it here as a derivative, the aggregate activity level is linear
in each $b_{i}$, and consequently this is the slope in $b_{i}$ for
any magnitude of change.) Players are more ``key'' if reducing their
$b_{i}$ causes a greater reduction in aggregate activity.\footnote{This is only one way of measuring what players are key, and in order for this to be a guide to interventions, various assumptions about costs and benefits of intervention have to hold. See \citet*{GGG} for more on this, and other network measures that might show up when we model the intervention differently.} It is a good exercise to show that the keyness vector can be expressed in terms
of Bonacich centrality. That, figure out how to fill in the question marks
in: 
\[
\bm{k}=\bm{\beta}(?;\alpha,?)
\]

\section{Limits in which long walks matter}

\subsection{The limit in which actions blow up}

Fix all parameters, in particular the vector $\bm{b}$, and consider
taking $\alpha\uparrow r(\bm{W})^{-1}$, so that $r(\alpha\bm{W})\uparrow1$.
What happens?

Well, let's stare at the sum 
\begin{equation}
\bm{a}^{*}=\sum_{\ell=0}^{\infty}\alpha^{\ell}\bm{W}^{\ell}\bm{b}.\label{eq:sum}
\end{equation}
 We know that for each $\alpha$ it is well-defined and finite. But
what happens to a typical entry in the limit?

As we saw in Problem Set 1, Problem 3, the maximum entry of $\bm{W}^{\ell}$
is asymptotically (as $\ell\to\infty$) exactly of order $r(\bm{W})^{\ell}$,
i.e. in the class $\Theta(r(\boldsymbol{W})^{\ell})$. From this we
can conclude that 
\begin{fact}
As $\alpha\uparrow r(\bm{W})^{-1}$, the sum (\ref{eq:sum}) tends
to $+\infty$. 
\end{fact}
Give an economic interpretation of this. (Basic idea: feedback effects
get out of control because we are weighting longer and longer walks.) 

\subsection{A coordination game and a related, but nicer, limit}

Take an irreducible, nonnegative, row-stochastic\footnote{A matrix is said to be row-stochastic if each of its rows adds up
to $1$.} matrix $\bm{\Gamma}$ with $\gamma_{ii}=0$ for each $i$ and set
$\bm{W=\bm{\Gamma}}$. Let $\bm{b}(\alpha)=(1-\alpha)\bm{y}$. Here
$\bm{y}$ is fixed. Note that for fixed parameters this is just a
special case of the general game we've been studying. However, now
that $\bm{b}$ depends on $\alpha$, the asymptotics of this model
in $\alpha$ will be different from the case in which $\bm{b}$ does
not depend on $\alpha$ from the previous subsection.

With the parameter values described above, the game we have been studying
is a coordination game: every player wants to match a weighted average
of (i) own ideal action $y_{i}$ and (ii) a weighted average of neighbors'
actions. You should check that in $i$'s best-response function
\begin{equation}
\text{BR}_{i}(\bm{a}_{-i})=\alpha\sum_{j}\gamma_{ij}a_{j}+(1-\alpha)y_{i}.\label{eq:BR-1}
\end{equation}
the weights placed on $y_{i}$ and the various $a_{j}$'s sum to $1$.

Applying Theorem \ref{thm:eqm-characterization} to characterize the
equilibrium, we find:
\begin{align}
\bm{a}^{*}=\sum_{\ell=1}^{\infty}\alpha^{\ell}\bm{\Gamma}^{\ell}\bm{b} & =(1-\alpha)\sum_{\ell=1}^{\infty}\alpha^{\ell}\bm{\Gamma}^{\ell}\bm{y}.\label{eq:sum-Gamma}
\end{align}
You'll show in Problem Set 2 that every player's action ends up being
an average of ideal points $y_{i}$ with certain weights.

Since it turns out that $r(\bm{\Gamma})=1$ (the spectral radius of
a row-stochastic matrix is equal to $1$) the characterization of
Theorem \ref{thm:eqm-characterization} holds for all $\alpha<1$.

Note that at $\alpha=1$ corresponds to a pure coordination game,
and you should convince yourself that the pure-strategy Nash equilibria
are exactly the action profiles with everyone taking the same action.
In particular, in the $\alpha=1$ game there is a huge amount of equilibrium
multiplicity.

For any $\alpha<1$, however, the game has a unique equilibrium. Also
in Problem Set 2, you'll verify that the equilibria as $\alpha\uparrow1$
converge to a well-defined limit in which all players take the same
action, no matter what that action is. Thus, we can think of the $\alpha\uparrow1$
limit as a way to refine the large set of equilibria in the $\alpha=1$
coordination game.

In the $\alpha\uparrow1$ limit, long walks (in $\bm{\Gamma}$) also
matter: this follows from (\ref{eq:BR-1}). But now they matter in
a limit that is better-behaved than the explosive limit studied in
the previous subsection. Here the long walks will determine the way
in which everyone averages the ideal points $y_{i}$ in setting their
actions.

\section{The Perron-Frobenius Theorem}

\subsection{Motivation}

As just discussed, we are interested in the $\alpha\uparrow r(\bm{W})^{-1}$
limit of our game, which will correspond to the type of Bonacich centrality
that 
\begin{itemize}
\item cares as much about network effects as possible subject to being well-defined
\item cares a lot about long walks.
\end{itemize}
The next result, which is a fundamental theorem that will recur repeatedly,
will help us think about this limit. Indeed, it will tell us essentially
everything about $\bm{W}^{\ell}$ for large $\ell$. But getting there
takes a little bit of setup.

\subsection{Perron-Frobenius Theorem}

This theorem goes a long way in the economic analysis of networks.\footnote{See \citet{elliott2018network-jpe} for a use of it in the context of characterizing efficient, rather than Nash equilibrium, outcomes.} A wonderful
reference on it is Carl D. Meyer's\emph{ \href{https://www.amazon.com/exec/obidos/ASIN/0898714540}{Matrix Analysis and Applied Linear Algebra, Chapter 8}};
I recommend this textbook very highly. A shorter self-contained exposition
can be found in {Debreu and Herstein's 1953 paper}
in \emph{Econometrica} \citeyearpar{debreu1953nonnegative}.

For any matrix $\bm{A}$, we denote by $\text{spec}(\bm{A})$ the
set of its eigenvalues. This set is also called its \emph{spectrum}.
\begin{defn*}
The \emph{spectral radius }of $\bm{A}$ is defined to be $$ r(\bm{A})=\max_{\lambda\in \operatorname{spec}(\bm{A})}| \lambda|,$$which
is the maximum absolute value of the eigenvalues of $\bm{A}$.
\end{defn*}
\begin{thm*}[Perron-Frobenius]
Let $\bm{A}$ be an irreducible, square matrix with no negative entries.
Then: 
\begin{enumerate}
\item The positive real number $r(\bm{A})$ is an eigenvalue (call it $\lambda_{1}$)
of $\bm{A}$. 
\item There is a unique vector\footnote{The notation $\mathbb{R}_{+}^{n}$ means the set of all vectors in
$\mathbb{R}^{n}$ with nonnegative entries. } $\bm{p}\in\mathbb{R}_{+}^{n}$ (called a right-hand Perron vector)
satisfying $\sum_{i}p_{i}=1$ and $\boldsymbol{Ap}=\lambda_{1}\boldsymbol{p}$.
All entries of this vector are strictly positive.
\item If there is a $\boldsymbol{v}\in\mathbb{R}_{+}^{n}\setminus\left\{ 0\right\} $
and $r'\in\mathbb{R}$ such that $\boldsymbol{Av=}r'\boldsymbol{v}$
then $\boldsymbol{v}$ is a positive scalar multiple of $\boldsymbol{p}$,
and $r'=\lambda_{1}$.
\end{enumerate}
\end{thm*}
We can apply the same result to $\bm{A}^{\intercal}$ to obtain a
unique vector $\bm{q}\in\mathbb{R}_{+}^{n}$ satisfying $\sum_{i}q_{i}=1$
and $\boldsymbol{A}^{\intercal}\bm{q}=r(\bm{A}^{\intercal})\boldsymbol{q}$. 

Because a matrix and its transpose have exactly the same eigenvalues,
the matrices $\bm{A}$ and $\bm{A}^{\intercal}$ have the same spectral
radius.\footnote{The fact that $r(\bm{A}^{\intercal})=r(\bm{A})$ is also easily deduced
from the solution to Problem Set 1, Problem 4, as long as we accept
that the definitions of $r(\bm{A})$ given there and in the present
note are equivalent.} Thus, taking the transpose of both sides of $\boldsymbol{A}^{\intercal}\bm{q}=r(\bm{A}^{\intercal})\boldsymbol{q}$,
we find that $\bm{q}^{\intercal}$ is a left-hand eigenvector of $\bm{A}$
satisfying $\bm{q}^{\intercal}\bm{A}=r(\bm{A})\boldsymbol{q}^{\intercal}$.
We will call $\bm{q}^{\intercal}$ the \emph{left-hand Perron vector}
of $\bm{A}$.

To summarize, the special (positive, real) eigenvalue $\lambda_{1}=r(\bm{A})$
of the matrix $\bm{A}$ is associated with two special eigenvectors
$\bm{p}$ (on the right) and $\bm{q}^{\intercal}$ (on the left),
each having only positive entries and each having entries summing
to $1$. The vector $\bm{p}$ is (up to normalization) the unique
nonnegative right-hand eigenvector of the matrix $\bm{A}$, and the
analogous statement holds for $\bm{q}^{\intercal}$ on the left-hand
side.

You have already met $r(\bm{A})$ from another angle, in Problem 4
of Problem Set 1, though this is the first time we are discussing
the Perron eigenvectors.

\subsection{Application to long walks}
\begin{prop}
Let $\bm{A}$ be an irreducible, square matrix with no negative entries.
Let $r(\bm{A})$ be the its largest eigenvalue ($\bm{A}$'s spectral
radius). As $\alpha\uparrow r(\bm{A})^{-1}$ we have 
\[
(1-\alpha r(\bm{A}))(\bm{I}-\alpha\bm{A})^{-1}\to\frac{\bm{p}\bm{q}^{\intercal}}{\bm{q}^{\intercal}\bm{p}}.
\]
\end{prop}
The right-hand side is a rank-$1$ matrix whose $(i,j)$ entry is
$cp_{i}q_{j}$, where the normalizing constant is the dot product
of $\bm{p}$ and $\bm{q}$. The proof of this is  left as an exercise.

\begin{defn}
A nonnegative matrix $\bm{A}$ is said to be \emph{primitive} if $\bm{A}^{\ell}$
has all strictly positive entries for some positive integer $\ell$. 
\end{defn}
\begin{prop}
The nonnegative, irreducible matrix $\bm{A}$ is primitive if and
only if

\[
\lim_{\ell\to\infty}\frac{\bm{A}^{\ell}}{r(\bm{A})^{\ell}}\to\frac{\bm{p}\bm{q}^{\intercal}}{\bm{q}^{\intercal}\bm{p}},
\]
where $\bm{p}$ and $\bm{q}^{\intercal}$ are the right-hand and left-hand
Perron vectors of $\bm{A},$ respectively.
\end{prop}
This is proved in \citep{meyer-book}.

\section{Comments and related models}

For our purposes, this resolves the question of how influential various individuals are on the group outcome. The results we just derived about the behavior of long walks are important when we study the long-run behavior of naive learning processes. In particular, the weight of long walks determines one's influence \citep{GolubJackson2010}. The \emph{rate} at which the approximations above become good is studied in \citep{DeMarzo2003,golub2012homophily}.

A very natural question is how to extend the analysis we have done to incomplete information. This is done in \cite*{Marti2015} \cite*{GolubMorris2017-ENC}, and \cite*{lambert_martini_ostrovsky}. \cite*{GolubMorris2016} shows that the linear algebra we have discussed above is closely related to \emph{higher-order expectations}, an important object in the study of beliefs and priors generally. 

\bibliographystyle{myecta}
\bibliography{enc_notes}

\begin{thebibliography}{16}
\newcommand{\enquote}[1]{``#1''}
\expandafter\ifx\csname natexlab\endcsname\relax\def\natexlab#1{#1}\fi

\bibitem[\protect\citeauthoryear{Acemoglu, Carvalho, Ozdaglar, and
  Tahbaz-Salehi}{Acemoglu et~al.}{2012}]{acemoglu2012network}
\textsc{Acemoglu, D., V.~M. Carvalho, A.~Ozdaglar, and A.~Tahbaz-Salehi}
  (2012): \enquote{The network origins of aggregate fluctuations,}
  \emph{Econometrica}, 80, 1977--2016.

\bibitem[\protect\citeauthoryear{Ballester, Calv\'{o}-Armengol, and
  Zenou}{Ballester et~al.}{2006}]{Ballester2006}
\textsc{Ballester, C., A.~Calv\'{o}-Armengol, and Y.~Zenou} (2006):
  \enquote{{Who's who in Networks. Wanted: the Key Player},}
  \emph{Econometrica}, 74, 1403--1417.

\bibitem[\protect\citeauthoryear{Baqaee and Farhi}{Baqaee and
  Farhi}{2017}]{baqaee2017macroeconomic}
\textsc{Baqaee, D.~R. and E.~Farhi} (2017): \enquote{The Macroeconomic Impact
  of Microeconomic Shocks: Beyond Hulten's Theorem,} Tech. rep., National
  Bureau of Economic Research.

\bibitem[\protect\citeauthoryear{Bonacich}{Bonacich}{1987}]{bonacich}
\textsc{Bonacich, P.} (1987): \enquote{Power and Centrality: A Family of
  Measures,} \emph{American Journal of Sociology}, 92, 1170--1182.

\bibitem[\protect\citeauthoryear{de~Mart{\'{\i}} and Zenou}{de~Mart{\'{\i}} and
  Zenou}{2015}]{Marti2015}
\textsc{de~Mart{\'{\i}}, J. and Y.~Zenou} (2015): \enquote{Network games with
  incomplete information,} \emph{Journal of Mathematical Economics}, 61,
  221--240.

\bibitem[\protect\citeauthoryear{Debreu and Herstein}{Debreu and
  Herstein}{1953}]{debreu1953nonnegative}
\textsc{Debreu, G. and I.~N. Herstein} (1953): \enquote{Nonnegative square
  matrices,} \emph{Econometrica: Journal of the Econometric Society}, 597--607.

\bibitem[\protect\citeauthoryear{DeMarzo, Vayanos, and Zwiebel}{DeMarzo
  et~al.}{2003}]{DeMarzo2003}
\textsc{DeMarzo, P.~M., D.~Vayanos, and J.~Zwiebel} (2003):
  \enquote{{Persuasion Bias, Social Influence, and Unidimensional Opinions},}
  \emph{The Quarterly Journal of Economics}, 118, 909--968.

\bibitem[\protect\citeauthoryear{Elliott and Golub}{Elliott and
  Golub}{2018}]{elliott2018network-jpe}
\textsc{Elliott, M. and B.~Golub} (2018): \enquote{A network approach to public
  goods,} \emph{Journal of Political Economy}, Forthcoming.

\bibitem[\protect\citeauthoryear{Galeotti, Golub, and Goyal}{Galeotti
  et~al.}{2017}]{GGG}
\textsc{Galeotti, A., B.~Golub, and S.~Goyal} (2017): \enquote{Targeting
  Interventions in Networks,} Available at SSRN:
  https://ssrn.com/abstract=3054353.

\bibitem[\protect\citeauthoryear{Golub and Jackson}{Golub and
  Jackson}{2010}]{GolubJackson2010}
\textsc{Golub, B. and M.~O. Jackson} (2010): \enquote{{Na{\"\i}ve Learning in
  Social Networks and the Wisdom of Crowds},} \emph{American Economic Journal:
  Microeconomics}, 2, 112--49.

\bibitem[\protect\citeauthoryear{Golub and Jackson}{Golub and
  Jackson}{2012}]{golub2012homophily}
\textsc{Golub, B. and M.~O. Jackson} (2012): \enquote{How homophily affects the
  speed of learning and best-response dynamics,} \emph{The Quarterly Journal of
  Economics}, 127, 1287--1338.

\bibitem[\protect\citeauthoryear{Golub and Morris}{Golub and
  Morris}{2017{\natexlab{a}}}]{GolubMorris2017-ENC}
\textsc{Golub, B. and S.~Morris} (2017{\natexlab{a}}): \enquote{Expectations,
  Networks and Conventions,} Available at SSRN:
  https://ssrn.com/abstract=2979086.

\bibitem[\protect\citeauthoryear{Golub and Morris}{Golub and
  Morris}{2017{\natexlab{b}}}]{GolubMorris2016}
\textsc{Golub, B. and S.~Morris} (2017{\natexlab{b}}): \enquote{Higher-Order
  Expectations,} Available at SSRN: http://ssrn.com/abstract=2979089.

\bibitem[\protect\citeauthoryear{Golub and Sadler}{Golub and
  Sadler}{2016}]{golub-sadler}
\textsc{Golub, B. and E.~Sadler} (2016): \enquote{Learning in Social Networks,}
  in \emph{The Oxford Handbook of the Economics of Networks}, ed. by
  Y.~Bramoull{\'e}, A.~Galeotti, B.~Rogers, and B.~Rogers, Oxford University
  Press, chap.~19, 504--542.

\bibitem[\protect\citeauthoryear{Lambert, Martini, and Ostrovsky}{Lambert
  et~al.}{2018}]{lambert_martini_ostrovsky}
\textsc{Lambert, N., G.~Martini, and M.~Ostrovsky} (2018): \enquote{Quadratic
  games,} Preprint, Stanford University,
  http://web.stanford.edu/~ost/papers/qg.pdf.

\bibitem[\protect\citeauthoryear{Meyer}{Meyer}{2000}]{meyer-book}
\textsc{Meyer, C.~D.} (2000): \emph{{Matrix Analysis and Applied Linear
  Algebra}}, Philadelphia: SIAM.

\end{thebibliography}

\end{document}